# Compact Modeling and Simulation of Heavy Ion Induced Soft Error Rate in Space Environment: Principles and Validation

Gennady I. Zebrev, Artur M. Galimov

*Abstract* — A simple physical model for calculation of the ion-induced soft error rate in space environment has been proposed, based on the phenomenological cross section notion. Proposed numerical procedure is adapted to the multiple cell upset characterization in highly scaled memories. Nonlocality of the ion impact has been revealed as the key concept determining the difference between physical processes in low scaled and highly scaled memories. The model has been validated by comparison between the simulation results and the literature on-board data. It was shown that proposed method provides single-valued prediction results correlating well with on-board data based solely on cross section data and LET spectra without any hidden fitting parameters and procedures.

## I. Introduction

THE problem of the single event effects in spaceborne microelectronics has been arisen and realized more than a three decades years ago [1]. The main concepts and principles of the soft error rate (SER) computation have been formulated at the same time [2]. Despite some modifications [3, 4, 5], the calculation methods developed at that time are still in use. Meanwhile, the memory cell sizes have been decreased over the years by several orders. As a result, we have at the moment a situation when the cross ion track sizes are larger than the memory cell sizes. Unlike the old low scaled circuits, in which not everyone heavy ion hit leads to a single bit error, the multiple cell upsets (MCU) are typical for modern highly scaled memories. The rise of multiple cell upsets has led to the new challenges for efforts to calculate the soft error rate (SER) in the space environment. As a matter of fact, the growing role of the MCUs questions the use of traditional SER calculation methods and software packages such as CREME96, SPENVIS, OSOT, etc. Most methods are based on the notion of a separate (isolated) sensitive volume (SV) having a shape of a rectangular parallelepiped (RPP) (see, e.g., Sec. 5.4 in [6]).

Such isolation of the single cell SV implicitly assumes the condition of the local impact of the ion when a single particle cannot strike more than one memory cell. The locality of the ion impact means that the cross area of the ion track is much smaller than the layout size of memory cells. Otherwise, it would be the case of the charge sharing between neighbor cell nodes. The ion impact locality is also a necessary condition for the cross section saturation at high LETs when any additional collected charge is unable to produce additional errors. The charge sharing violates the locality condition in the highly scaled integrated circuits. This violation manifests itself experimentally in the two facts: (i) the dominance of the MCUs in highly scaled memory IC; (ii) a lack of cross section saturation at high LETs. Furthermore, the notion of the isolated RPP SV together with its formally accurate chord length distributions does also breakdown [1]. Novel simple and physically consistent approaches to the soft error rate calculation are needed. The aim of the present paper is to describe a compact model for the SER calculation, allowing the SER prediction in a simple, straightforward and single-valued manner.

The rest of this paper is organized as follows. First, we discuss in Sec. II the uncertainties of the traditional approach based on the cross section curve Weibull interpolation. Sec. III is devoted to a description of proposed SER computational procedure. The model validation is presented and discussed in Sec. IV and V.

## II. Uncertainties of the Weibull parameter extraction

The standard SEE sensitivity characterization methods rely, in particular, on the interpolation of the average single bit upset cross section dependence on LET (denoting here as $\Lambda$) in the form of the 4-parameter Weibull curve

$$\sigma(\Lambda) = \sigma_{SAT}\left(1 - \exp\left[-\left(\frac{\Lambda - \Lambda_C}{W}\right)^s\right]\right). \quad (1)$$

The Weibull function has the two principal parameters, namely, the saturation cross section $\sigma_{SAT}$ and the threshold (critical) value $\Lambda_C$ of LET. Two additional adjusting parameters ($W$ and $s$) determine the shape of the cross section curve in a sub-threshold region and a transition region between the sub-threshold and the above-threshold parts of the cross section dependencies. Despite the common practice, a use of the Weibull interpolation encounters the challenges related to the ambiguity of the basic parameters and extraction procedures [7]. In particular, the Weibull curve is a function having a distinct saturation, which a priori implies an existence of the experimental cross section saturation at high LETs. Meanwhile, the lack of saturation in the cross sections up to 120 MeV-cm$^2$/mg has been reported by many investigators [8, 9,10,11,12]. The ambiguity of the Weibull function can be illustrated by the following example. Figure 1 shows a comparison of the two Weibull approximation of the same experimental data.

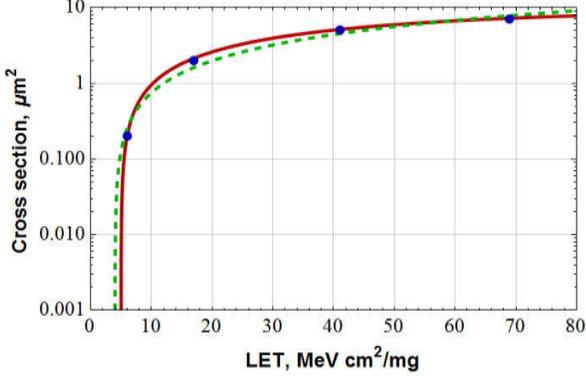

Fig.1. Two Weibull functions, describing a single set of experimental points with two different parameter sets. Solid line: $\sigma_{SAT}$ = 10 μm$^2$, $\Lambda_C$ = 5 MeV×cm$^2$/mg, W = 50 MeV×cm$^2$/mg, s =1. Dashed line: $\sigma_{SAT}$ = 100 μm$^2$, $\Lambda_C$ = 4 MeV×cm$^2$/mg, W =800 MeV×cm$^2$/mg, s = 1.

As can be seen in Fig. 1 which is taken from [13], the same set of experimental points can be satisfactorily approximated by the Weibull curves with drastically different parameters. Due to parametric over-determination of the Weibull function, any uncertainty in the saturation cross section $\sigma_{SAT}$ can be compensated by changing of other parameters, in particular, $W$. Thus, the numerical value of $\sigma_{SAT}$ without specifying the additional parameter $W$ is not a good informative parameter.

Any spread in the input parameters results in the spread in the calculated error rates. For example, employing two set of the "equivalent" Weibull parameters in Fig. 1 with a modified figure-of-merit expression [14]

$$R = \frac{0.5 b \sigma_{SAT}}{\left(\Lambda_C + W \times 0.288^{1/s}\right)^2}, \qquad (2)$$

we have obtained a spread of the SER estimations more than an order of magnitude. Since uncertainties in $\sigma_{SAT}$ inevitably leads to uncertainties of the lateral RPP sizes, the similar problems immediately arise also in computation with the RPP based CREME96, SPENVIS, OSOT, etc.

III. SOFT ERROR RATE CALCULATION

The soft error rate can be simulated without using the poorly defined parameters of isolated RPP sensitive volume for individual memory cells. Soft error rate can be generally calculated based immediately on definition of the phenomenological cross section notion as an integral convolution

$$R = M \int_0^\infty \sigma(\Lambda) \phi(\Lambda) d\Lambda, \qquad (3)$$

where $M$ is the total memory cell number, $\sigma(\Lambda)$ is the LET-dependent cross section of single bit upsets (SBU) per a bit, averaged over the full solid angle [15], $\phi(\Lambda)$ is the heavy ion LET omnidirectional spectrum. In fact, the phenomenological cross section notion, underlying the general formula (3), is experimentally determined as the response of the whole circuit followed by a formal (and logically optional) normalization to a single cell.

A. Petersen's FOM approach

The formula (3) generally requires a numerical integration. Therefore, some simplified methods are often used in practice, e.g., the figure-of-merit approach, proposed originally by Petersen et al. in 1983 [16]. The original Petersen approach is mainly based upon the two assumptions:
1. The cross section vs LET dependence is approximated through the step function $\theta(x)$ as follows

$$\sigma(\Lambda) = \sigma_{SAT} \theta(\Lambda - \Lambda_C). \qquad (4)$$

2. The differential LET spectrum is assumed to be a power function in the range 2 - 20 MeV-cm$^2$/mg

$$\phi(\Lambda) \cong b/\Lambda^3, \qquad (5)$$

where $b$ is an orbit-dependent constant. Then the soft error rate per bit can be estimated as follows

$$R_{Petersen} = \int_0^\infty \sigma(\Lambda) \phi(\Lambda) d\Lambda \cong b\sigma_{SAT} \int_{\Lambda_C}^\infty \Lambda^{-3} d\Lambda =$$
$$= 0.5\sigma_{SAT} \, b/\Lambda_C^2 = \sigma_{SAT} \phi(\Lambda > \Lambda_C), \qquad (6)$$

where $\phi(\Lambda > \Lambda_C) = 0.5b/\Lambda_C^2$ is an integral ion flux for LETs, exceeding the critical value $\Lambda_C$. Equation (2) is a later modification of (6). As noticed above, (2) and (6) contain ambiguous parameters.

B. Alternative data parametrization

As an alternative approach, another analytic form for experimental data interpolation has been proposed in [17]

$$\sigma(\Lambda) = K_d W \ln\left[1 + \exp\left(\frac{\Lambda - \Lambda_C}{W}\right)\right], \qquad (7)$$

where $K_d$ is a differential slope per bit of the above-threshold cross section curve ($\Lambda > \Lambda_C$), $W$ is a parameter responsible for the cross section behavior in the subthreshold region ($\Lambda < \Lambda_C$). The subthreshold region gives rather a small contribution to the SER for the highly scaled memories with low $\Lambda_C$. Then, the cross section vs. LET dependence above the threshold ($\Lambda > \Lambda_C$) can be approximated by a simple linear form

$$\sigma(\Lambda) \cong K_d (\Lambda - \Lambda_C), \qquad \Lambda > \Lambda_C, \qquad (8)$$

where the slope $K_d$ is used instead of $\sigma_{SAT}$ for parametrization of the above-threshold cross section curves.

In contrast to the ill-defined saturation value $\sigma_{SAT}$, the slope $K_d$ can be obtained uniquely, for example, by the least squares method. Particularly, Figs. 2 show the test results obtained for the same 4 Mb 90 nm device in the different LET ranges [13].

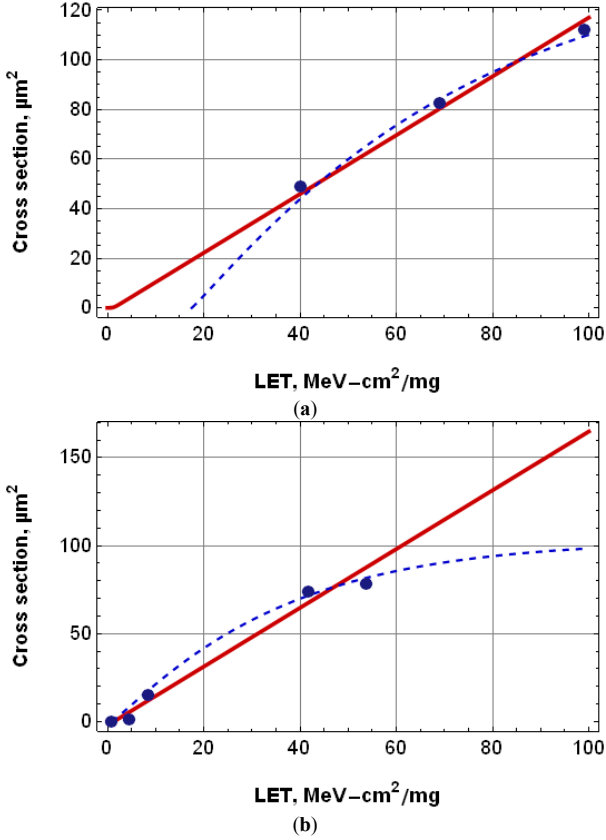

Fig. 2. Cross section data of the same 90 nm memory measured for two type of experimental facilities with different LET ranges. Both sets of experimental data were interpolated with the Weibull (dashed lines) and the quasi-linear (solid lines) functions. Fitted parameters are shown in Table I.

Figure 2 illustrates the following disadvantages of the Weibull approximations: (i) the sensitivity of extracted $\sigma_{SAT}$ to the upper LET values; (ii) the sensitivity of extracted $\Lambda_C$ to the lower LET values; (iii) a lack of a unique way to fit the Weibull parameters. As a consequence, the SER calculated with parameters, fitted from the data in (2a), turns out to be 6 times greater than for the data in (2b).

TABLE I. COMPARISON OF WEIBULL AND LINEAR FITTED PARAMETERS

| Appr | Weibull | | | | Linear | |
|---|---|---|---|---|---|---|
| Params | $\sigma_{SAT}$, μm² | $\Lambda_C$, MeV-cm²/mg | $W$, MeV-cm²/mg | $s$ | $\Lambda_C$, MeV-cm²/mg | $K_d$, mg/MeV |
| (a) | 143 | 17 | 57 | 1.07 | 1.2 | 1.2×10⁻⁸ |
| (b) | 103 | 1 | 35 | 1.09 | 1.3 | 1.7×10⁻⁸ |

In contrast to the Weibull function, the linear interpolation turned out to be more robust (see Table I). A slightly decreased value $K_d$ at high LETs can be explained by the charge yield reducing due to the Auger recombination [18].

### C. Alternative FOM approach

Based on the same approximation for the LET spectrum (5) together with (8) and (3), one can easily derive another form of figure-of-merit (FOM) relation

$$R = M\int_0^\infty \sigma(\Lambda)\phi(\Lambda)d\Lambda \cong bM \int_{\Lambda_C}^\infty K_d(\Lambda - \Lambda_C)\Lambda^{-3}d\Lambda = \\ = b\frac{MK_d}{2\Lambda_C} = b\frac{MK_d\Lambda_C}{2\Lambda_C^2} = MK_d\Lambda_C\phi(\Lambda > \Lambda_C). \quad (9)$$

The ratio of (9) to the Petersen expression (6) can be estimated as follows

$$\frac{R}{R_{Petersen}} = \frac{K_d\Lambda_C}{\sigma_{SAT}} \cong \frac{\Lambda_C}{\Lambda_{max} - \Lambda_C}. \quad (10)$$

It is assumed here that $\sigma_{SAT}$ is usually estimated at a maximum value of LET $\Lambda_{max}$, which corresponds to $\sigma_{SAT} \cong MK_d(\Lambda_{max} - \Lambda_C)$. This means that for $\Lambda_{max} > 2\Lambda_C$ (typical for modern commercial devices), the Petersen FOM provides a more conservative estimate. Importantly, both types of FOM are inappropriate for the low critical LETs, typical for the modern highly scaled memories.

### D. Generalized compact SER model

Equations (3) and (8) allows calculating the average bit-flip number $N_{SBU}$ as follows

$$N_{SBU} = M\int \sigma(\Lambda)\Phi(\Lambda)d\Lambda \cong MK_d \int_{\Lambda_C}(\Lambda - \Lambda_C)\Phi(\Lambda)d\Lambda = \\ = MK_d[\langle\Lambda\rangle - \Lambda_C]\Phi(\Lambda > \Lambda_C), \quad (11)$$

where $\Phi(\Lambda)$ is the differential LET spectrum for omnidirectional fluence, $\langle\Lambda\rangle$ is an effective LET, averaged over of the "hard" (i.e., $\Lambda \geq \Lambda_C$) part of the LET spectrum

$$\langle\Lambda\rangle = \frac{\int_{\Lambda_C}^\infty \Lambda\Phi(\Lambda)d\Lambda}{\int_{\Lambda_C}^\infty \Phi(\Lambda)d\Lambda}. \quad (12)$$

By its definition, the condition $\langle\Lambda\rangle > \Lambda_C$ is always satisfied. The effective LET or, the average $\langle\Lambda\rangle$, defined in (12), is determined in the main by the orbit parameters with a specific LET spectrum, and, partly, by the circuit properties through dependence on $\Lambda_C$. It is, loosely speaking, a "center of mass" of a LET distribution, defined for $\Lambda > \Lambda_C$. A graphical illustration of SER computation procedure is shown in Fig. 3.

As can be seen in Fig. 3, the main contribution to the integral SER in highly scaled circuits comes from the region around $\langle\Lambda\rangle$. This means that the contribution of high LETs to the integral SER turns out to be often insignificant in this case. In other words, the cross section dependence interpolation accuracy at the high LETs becomes not critical. This is another argument in favor of the linear cross section interpolation near $\langle\Lambda\rangle$ despite a tendency to a sub-linear behavior, which is typically observed at high LETs. Such approach is certainly justified for the highly scaled commercial memories, wherein $\Lambda_C$ and $\langle\Lambda\rangle$ are much less than the maximal LETs $\Lambda_{max}$ under tests.

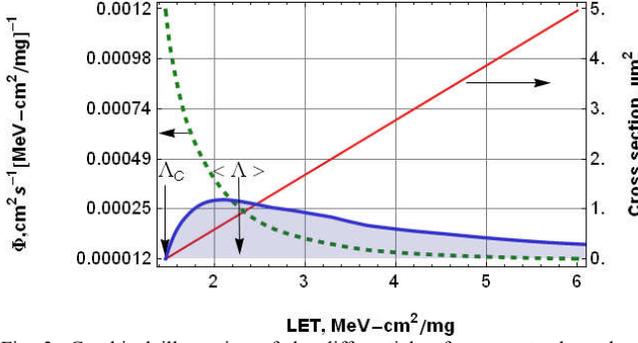

Fig. 3. Graphical illustration of the differential soft error rate dependence (thick solid curve) as a convolution of the quasi-linear cross section dependence on LET (solid line) and the LET spectrum (dashed line). According to (3), the integral SER is graphically represented here by the square of the shaded region [13].

In practice, (11) implies the numerical calculations based on a known LET spectrum at a given space orbit. We have developed a simple software application (PRIVET II), which allows quick calculations of SER based solely on the differential LET spectrum in a standard table form and on several experimental cross section points.

### E. Linear interpolations at different tilt angles

It is well known that the upset cross section data are generally dependent on the tilt angle $\theta$ of the incident ion. For example, the cross section tilt dependencies for the HM628512 and M5M5408 SRAMs have been experimentally examined by Petit et al. [8]. We interpolated these curves by the least squares method (see Figs. 4 and 5).

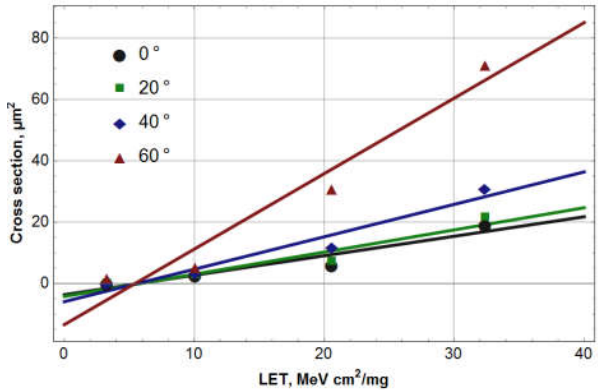

Fig. 4. The least square approximations of the HM628512 SRAM cross section data (adapted from [8]) at the different ion incident tilts $\theta$ = 0°; 20°; 40°; 60°. Fitted parameters are shown in Table II.

TABLE II. PARAMETERS FITTED FOR DATA HM628512 IN FIG. 4.

| Polar angle $\theta$ | 0° | 20° | 40° | 60° |
|---|---|---|---|---|
| $K_d$, $10^{-9}$ mg/MeV | 6.35 | 7.22 | 10.5 | 24.6 |
| $\Lambda_C$, MeV-cm$^2$/mg | 5.7 | 5.78 | 5.57 | 5.43 |

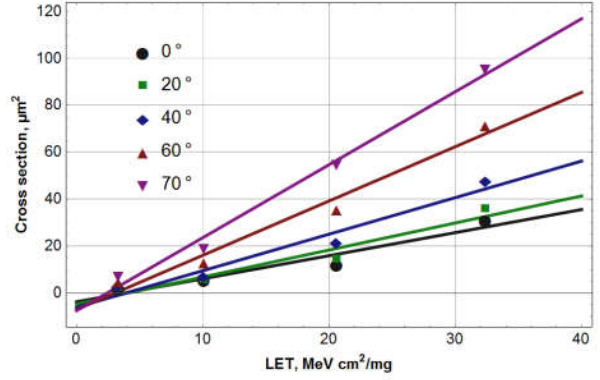

Fig. 5. Linear approximations for the M5M5408 SRAM experimental data at different polar angles $\theta$ = 0°; 20°; 40°; 60°; 70°. Fitted parameters are shown in Table III.

TABLE III. PARAMETERS FITTED FOR DATA M5M5408 IN FIG. 5.

| Polar angle $\theta$ | 0° | 20° | 40° | 60° | 70° |
|---|---|---|---|---|---|
| $K_d$, $10^{-9}$ mg/MeV | 9.82 | 11.5 | 15.5 | 23.1 | 31.0 |
| $\Lambda_C$, MeV-cm$^2$/mg | 3.80 | 4.07 | 3.88 | 3.04 | 2.41 |

As can be seen in Figs. 4-5, the cross section curves do not saturate at high LETs and can be well approximated with linear dependence for all tilt angles. With the growing of the tilt angle, the slope $K_d$ also increases. This behavior will be discussed below in the Sec. V.

## IV. MODEL VALIDATION

### A. SAC-C Mission Data Analysis

The proposed compact model has been validated with a comparison between the simulation results and in-flight data based on the detailed results provided by Boatella et al. in 2010 [19] and Falguere et al. in 2002 [20]. Particularly, we have compared in this section the SAC-C mission in-flight SER data [19] with our calculation results for the HM628512 and the KM684000 devices. The ground test data for heavy ion cross section, provided in [20], were used to extract $K_d$ and $\Lambda_C$ parameters by the ordinary least squares method (see Fig. 6).

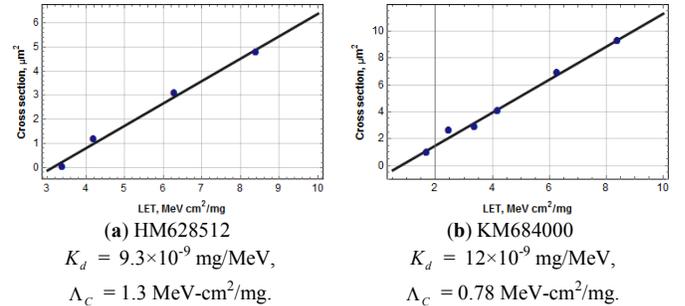

(a) HM628512
$K_d$ = 9.3×10$^{-9}$ mg/MeV,
$\Lambda_C$ = 1.3 MeV-cm$^2$/mg.

(b) KM684000
$K_d$ = 12×10$^{-9}$ mg/MeV,
$\Lambda_C$ = 0.78 MeV-cm$^2$/mg.

Fig. 6. The linear characterization and fitting parameters for heavy ion cross section data for two types of 4 Mbit SRAMs: (a) HM628512 and (b) the KM684000.

We have been used the OMERE tool [21] to calculate the heavy ion LET spectra for several years (2001-2008) with the following parameters: the SAC-C satellite orbit (707 km, 98.2°), M = 1 parameter for the GCR spectra, solar conditions and shielding (30 mm Al) were chosen the same as during the SAC-C mission [19].

We have numerically convolved the LET spectra with the linearly interpolated cross sections and compared obtained SER with on-board data for GCR (see, [19] Table II, the bottom cells). Comparisons of on-board data and our simulations are shown in Fig. 7(a) and (b).

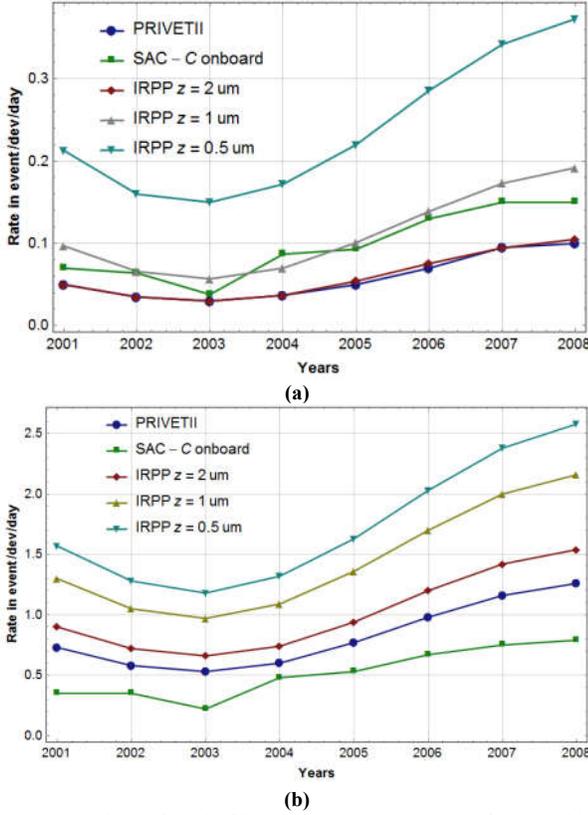

Fig. 7. Comparison of the in-flight and calculated SER, performed with our approach and with CREME96 for two type of SRAM: (a) HM628512 and (b) KM684000.

As can be seen in Fig. 7, the simulation results are in good agreement with in-flight data. At the same time, the CREME96 calculations show a wide spread in the results depending on the choice of the sensitive volume depth Z.

### B. Proba-II mission's in-flight data analysis

The proposed approach has been validated with a comparison between the calculation results and the in-flight data reported by M. D'Alessio et al. [22]. Particularly, in this section, the PROBA-II mission in-flight SER data have been compared with our calculation results for the AT68166 Multi-Chip module built with four 0.25 μm AT60142F 4 Mbit devices. The ground test data, provided for these circuits by Harboe-Sørensen et al. [11], have been used to extract $K_d$ and $\Lambda_C$ parameters by the ordinary least squares method. Figure 8 shows the experimental data in the logarithmic and linear plot. Remarkably, the experimental curve is close to linear up to 110 MeV-cm$^2$/mg.

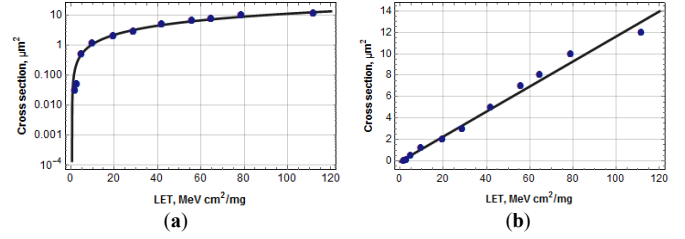

Fig. 8. Atmel AT60142F SRAM cross section data, depicted in logarithmic (a) and linear (b) scales. Parameters of the linear approximation are as follows: $K_d = 1.17 \times 10^{-9}$ mg/MeV, $\Lambda_C = 0.5$ MeV-cm$^2$/mg.

The LET spectrum has been generated in the OMERE tool with the PROBA-II orbit parameters as in [22]: 700 km altitude, 98.28° inclination, 3 years duration since March 1, 2010, 10 mm aluminum shield and CRÈME96 solar maximum conditions. The results of the comparison in Table IV show that despite the fact that our method has no free fitting parameters, it gives the best agreement with the on-board data compared to the IRPP calculations.

TABLE IV. COMPARISON OF ACTUAL AND CALCULATED SER FOR THE 16 MBIT AT68166 MODULE [11, 22].

| Actual in-flight SER | 0.168 upsets/ day |
|---|---|
| Our method | 0.132 upsets/ day |
| IRPP; Z =2 μm | 0.01 upsets/ day |
| IRPP; Z = 1 μm | 0.03 upsets/ day |
| IRPP; Z = 0.5 μm | 0.035 upsets/ day |

Note that a decreasing dependence of SER upon the SV depth Z, typical for IRPP method, is a controversial artifact of traditional methods.

### C. RHBD SRAM On-Orbit Data Analysis

The computational procedure described in Sec. III B is not quite general since it is adapted to the commercial circuits with the low critical LETs. The radiation-hardened memories are typically characterized by relatively high critical LETs when the contribution to SER from the subthreshold region $\Lambda < \Lambda_C$ must be accurately estimated. Particularly, such data with a pronounced subthreshold region in cross sections for 4 Mbit 0.25 μm RHBD SRAM's were presented in [23]. We have used in this case the full "logarithmic" form of the cross section approximation (7) to accurately interpolate the test data both in the above threshold and in the subthreshold regions. Fig. 9 shows the interpolation results depicted on logarithmic and linear scales. Notice that the subthreshold part of the data is close to exponential, which allows us to use the single-valued least square procedure to extract $W$ in the logarithmic scale.

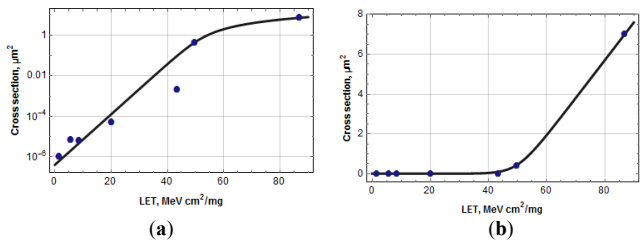

Fig. 9. Approximation of the cross section data (adapted from [23]) with equation (7), depicted in logarithmic (a) and linear (b) scales. Parameters are as follows: $K_d = 1.9\times10^{-9}$ mg/MeV, $\Lambda_C = 50$ MeV-cm$^2$/mg, $W = 3.5$ MeV-cm$^2$/mg.

Bogorad et al. [24] provided the results of the SER calculations and some on-orbit SER data for the same SRAMs.

TABLE V. SUMMARY OF UPSET RATE COMPARISON.

| Environment/Model description | RPP $Z = 2.25$ μm (errors/bit/day) | RPP $Z = 0.25$ μm (errors/bit/day) | Compact model (errors/bit/day) |
|---|---|---|---|
| CRÈME 96 Solar Min | $3.57\times10^{-11}$ | $6.06\times10^{-8}$ | $1.36\times10^{-9}$ |
| CRÈME 96 Solar Max | $4.50\times10^{-12}$ | $8.78\times10^{-9}$ | $4.25\times10^{-10}$ |
| CRÈME 86 M=1 | $1.80\times10^{-11}$ | $2.43\times10^{-8}$ | $1.06\times10^{-9}$ |
| In-flight data [24] | $1.8\times10^{-10}$ errors/bit/day | | |

A comparison between the IRPP calculations [24] and our compact simulations with the parameters in Fig. 9 is presented in a Table V. Despite we used the same LET spectra, our calculation scheme provides the unambiguous results that do not depend on the choice of the ill-defined model parameters. Notice that our SER predictions are rather close to on-orbit $1.8\times10^{-10}$ errors/bit/day, indicated in [24]. We have found that the use of logarithmic interpolation is a critical point since the neglect of the region $\Lambda < \Lambda_C$ leads to a strong underestimate in SER. This point is a common feature of the rad-hard memories with the high threshold LETs. Unlike the non-hardened high density ICs with low $\Lambda_C$, in which the SER contribution from the subthreshold ($\Lambda < \Lambda_C$) region is often insignificant, the SER in the rad-hard ICs with high $\Lambda_C$ can be very sensitive to the cross section magnitudes in the "shallow" subthreshold region $\Lambda \leq \Lambda_C$ and rather insensitive to the above threshold ($\Lambda > \Lambda_C$) part of the dependence due to extremely low ion fluxes at high LETs.

## V. DISCUSSION

Despite the use of different sets of parameters and computation procedures, the proposed approach, as well as other traditional methods (CREME96, SPENVIS, OSOT, etc.), is based on the microdosimetric approach. The main conceptual difference with the traditional methods is the following. Unlike the traditional methods, which are based on the response of independent, isolated and identical RPP sensitive volumes with the RPP chord length distribution, we, in fact, employ the notion of a single sensitive volume in a form of thin lamina. The validity of this notion is justified by nonlocality of the ionizing ion impact. The ion impact nonlocality effects include, particularly, the MCUs, which are most clearly pronounced in the highly scaled memories, as illustrated in Fig. 10.

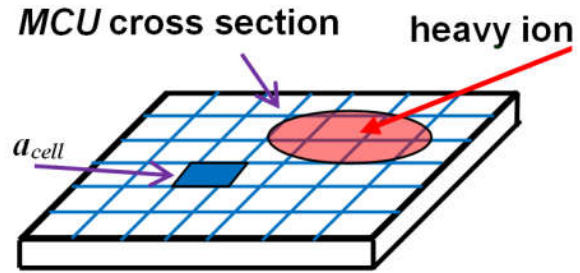

Fig. 10. Illustration of nonlocal nature of the ion impact in a highly scaled memory [13].

In fact, any form of charge sharing, occurring even in the low scaled circuits can be considered as the nonlocal effects. The radiation response of the entire memory circuit can not be calculated in the presence of non-local effects as the sum of the responses of individual memory cells since the chord length distribution for the isolated volumes becomes meaningless in this case.

At the same time, the impact nonlocality transforms a set of the isolated SVs of separate cells into the sensitive volume of the memory as a single whole. The area of this single SV is well-determined by the total memory area $M a_{cell}$, where $a_{cell}$ is the memory layout cell area.

It is important, that the problem of the SV depth determination, critical for the RPP-based methods, is completely absent in our approach. In fact, information about of the SV depth is implicitly contained in the experimentally determined slope $K_d$. The slope of cross section curve $K_d$ has the dimension of the inverse absorbed dose (1 mg/MeV = 62,415 rad$^{-1}$, 1 MeV/mg = $1.602\times10^{-5}$ rad) and a meaning of the coefficient of the quasi-linear dependence of SBU total number on LET [25]. Such a quasi-dose effect is a direct consequence of non-local nature of the ion impact in nanoscale memories when the charge collection is averaged over several memory cells (not necessarily adjacent) covered by a single ion track. In the simplest dose model, the slope $K_d$ per a bit can be estimated as follows [18, 25]

$$K_d = a_{cell} \frac{\rho t_{eff}}{\varepsilon_C}, \qquad (13)$$

where $\rho$ is the Si mass density, $\varepsilon_C$ is the memory cell critical energy, $t_{eff}$ is the effective thickness of the SV, including in itself the efficiencies of the charge generation, charge yield, and collection. The Poisson event scatter influence on the accuracy of $K_d$ extraction can be reduced with the statistical analysis methods [26].

As we have noted above, the slopes $K_d$ and, correspondingly, the effective SV depths $t_{eff}$ are, of course, the ion incidence angle dependent values [8]. To correctly model the omnidirectional ion flux of space environment, the test should be carried out at the incidence angle equal to 60 degrees. This incidence angle corresponds exactly to the SV average chord length, which is equal to twice the thickness of the SV for a thin lamina case.

The contribution to the SER of nuclear reactions below the threshold LETs makes the problem more difficult, necessitating the use of the Monte Carlo simulation-based approach [23, 27, 28]. Monte Carlo methods have also been used for the description of deviations from the isolated SV approximation [29]. Nevertheless, the numerical Monte Carlo simulations are rather computationally intensive and inconvenient in practical applications due to lack of detailed information about layout and composition of the ICs. Nonlocality of nuclear interactions makes it possible to neglect a fine structure of the charge collection regions, replacing them by a homogeneous single sensitive volume. Thus, the nonlocal impact allows us to significantly simplify the use of Monte Carlo calculations for estimation of average energy deposition in the single circuit SV. For example, the Monte Carlo techniques could provide the LET spectra of secondary products of nuclear reactions $\Phi_{SEC}(\Lambda)$, taking into account the average chemical composition of high-Z materials in the thick overlayers of modern ICs. Then, the contribution of nuclear reactions to SER in the circuits, that have a relatively high direct ionization threshold LET [30], could be calculated with equations similar to (11) and (12), based on the data for heavy ion direct ionization. This subject requires further investigations.

## VI. Summary

It has been shown that the use of the Weibull function can lead to a non-unique determination of its parameters, which, in turn, may cause the noticeable uncertainty in SER calculations. It is argued in this work that this mathematical confusion stems from the non-local character of the ion impact, which could lead to a lack of cross section saturation even at high LETs. The "logarithmic" interpolation function and parameters provide more flexible and one-valued fitting of the SEU cross section both in above threshold and subthreshold regions of the dependence on LET. We argue here that the basic concept of the IRRP method, such as the isolated sensitive volume with a well-defined chord length distribution for a separate memory cell, is also violated due to the particle impact nonlocality, especially pronounced in the highly scaled circuits.

We have been proposed and validated a simple compact model for fast and unique SER estimation, which has a form of a generalized figure-of-merit method. The main advantage of our approach lies in its straightforward, simple, single-valued calculation method based on a consistent physical picture. This method is purely phenomenological, and it is based solely on the experimental cross section data and on the LET spectra without any additional information about the physical mechanisms of nuclear interactions, circuit response, charge transport and collection, etc. The results of such very complicated processes are assumed to be included in the observed cross section. The accuracy of our method does not depend on any additional assumptions, and it is determined solely by the accuracy of the orbit LET spectrum and by the phenomenological cross section measurement and interpolation.

REFERENCES

[1] E. L. Petersen. R. Koga, M. A. Shoga, J. C. Pickel, W. E. Price, "Single Event Revolution," *IEEE Trans. Nucl. Sci.*, Vol. 60, No. 3, pp. 1824-1825, 2013.

[2] J. C. Pickel and J. T. Blandford, Jr., "CMOS RAM cosmic-ray-induced-error-rate analysis," *IEEE Trans. Nucl. Sci.*, Vol. 6, pp. 3962–3967, Dec. 1981.

[3] A. J. Tylka, J. H. Adams, P. R. Boberg et al. "CREME96: a revision of the Cosmic Ray Effects on Micro-Electronics code," *IEEE Trans. Nucl. Sci.*, Vol. 44, No. 6. pp. 2150–2160, Dec. 1997.

[4] J. H. Adams, A. F. Barghouty, M. H. Mendenhall, R. A. Reed et al. "CRÈME: The 2011 Revision of the Cosmic Ray Effects on Micro-Electronics Code," *IEEE Trans. Nucl. Sci.*, Vol. 59, No. 6, pp. 3141–3147, Dec. 2012.

[5] G. I. Zebrev, I. A. Ladanov et al., "PRIVET - Heavy Ion Induced Single Event Upset Rate Simulator in Space Environment," *RADECS 2005 Proceedings*, pp. 131-134, 2005.

[6] G. C. Messenger, M. Ash, *Single Event Phenomena*, Springer-Science. Chapman & Hall, 1997.

[7] Edward Petersen, *Single Event Effects in Aerospace*, John Wiley & Sons, Inc., Hoboken, New Jersey, USA, 2011.

[8] S. Petit, J. P. David, J. Falguere, S. Duzellier, C. Inguimbert, T. Nuns, and R. Ecoffet, "Memories response to MBU and semi-empirical approach for SEE rate calculation," *IEEE Trans. Nucl. Sci.*, Vol. 53, No. 4, pp. 1787 – 1793, Aug. 2006.

[9] D. F. Heidel, P. W. Marshall, J. A. Pellish et al., "Single-Event Upsets and Multiple-Bit Upsets on a 45 nm SOI SRAM," *IEEE Trans. Nucl. Sci.*, Vol. 56, No. 6, pp. 3499–3504, Dec. 2009.

[10] E. L. Petersen, "Rate Predictions for Single-Event Effects - Critique II," *IEEE Trans. Nucl. Sci.*, Vol. 52, No. 6, pp. 2158–2167, Dec. 2005.

[11] R. Harboe-Sørensen, C. Poivey, F.-X. Guerre, A. Roseng, F. Lochon, G. Berger, W. Hajdas, A. Virtanen, H. Kettunen, and S. Duzellier, "From the reference SEU Monitor to the Technology Demonstration Module On-Board PROBA-II," *IEEE Trans. Nucl. Sci.*, Vol. 55, No. 6, pp. 3082–3087, Dec. 2008.

[12] R. A. Weller, R. A. Reed, K. M. Warren, M. H. Mendenhall, B. D. Sierawski, R. D. Schrimpf, and L. W. Massengill, "General framework for single event effects rate prediction in microelectronics," *IEEE Trans. Nucl. Sci.*, Vol. 56, No. 6, pp. 3098–3108, Dec. 2009.

[13] V. S. Anashin, A. E. Koziukov, K. S. Zemtsov, M. E. Gorchichko, A. M. Galimov, G. I. Zebrev, "Compact Modeling of Soft Error Rate in Space Environment," to be published in *RADECS-2016 Proceeding*, 2016.

[14] E. L. Petersen, "The SEU Figure of Merit and Proton Upset Rate Calculation," *IEEE Trans. Nucl. Sci.*, Vol. 45, No. 6, pp. 2550-2562, Dec. 1998.

[15] G. I. Zebrev, I. O. Ishutin, R. G. Useinov, V. S. Anashin, "Methodology of Soft Error Rate Computation in Modern Microelectronics," *IEEE Trans. Nucl. Sci.,* Vol. 57, No.6, pp. 3725-3733, Dec. 2010.

[16] E. L. Petersen, J. B. Langworthy, and S. E. Diehl, "Suggested Single Event Upset Figure of Merit," *IEEE Trans. Nucl. Sci.,* Vol. 30, No. 6, pp. 4533-4539, Dec. 1983.

[17] G. I. Zebrev, M. S. Gorbunov, R. G. Useinov, V. V. Emeliyanov, A. I. Ozerov, V. S. Anashin, A. E. Koziukov, K. S. Zemtsov, "Statistics and methodology of multiple cell upset characterization under heavy ion irradiation," *Nuclear Instruments and Methods in Physics Research Sec. A*, Vol. 775, pp. 41-45, March 2015.

[18] G. I. Zebrev, K. S. Zemtsov, "Multiple cell upset cross section modeling: A possible interpretation for the role of the ion energy-loss straggling and Auger recombination," *Nuclear Instruments and Methods in Physics Research Sec. A*, Vol. 827, pp. 1-7, Aug. 2016.

[19] C. Boatella, G. Hubert, R. Ecoffet, and F. Bezerra, "ICARE on-board SAC-C: More than 8 years of SEU & MBU, analysis and prediction," *IEEE Trans. Nucl. Sci.*, Vol. 57, No. 4, pp. 2000-2009, Aug. 2010.

[20] D. Falguere, D. Boscher, T. Nuns, S. Duzellier, S. Bourdarie, R.Ecoffet, S. Barde, J. Cueto, C. Alonzo, and C. Hoffman, "In-flight observations of the radiation environment and its effects on devices in the SAC-C polar orbit," *IEEE Trans. Nucl. Sci.*, Vol. 49, No. 6, pp. 2782–2787, Dec. 2002.

[21] Outil de Modélisation de l'Environnement Radiatif Externe OMERE [Online]. Available: http://www.trad.frOMERE-Software.html

[22] M. D'Alessio, C. Poivey, V. Ferlet-Cavrois, P. Matthijs, "SRAMs SEL and SEU In-flight Data from PROBA-II Spacecraft," *RADECS-2013 Proceedings*, 2013.

[23] K. M. Warren, R. A. Weller, M. H. Mendenhall, R. A. Reed, D. Ball, C. Howe, B. Olson, M. Alles, L. Massengill, R. D. Schrimpf, N. Haddad, S. Doyle, D. McMorrow, J. Melinger, and W. Lotshaw, "The contribution of nuclear reactions to heavy ion single event upset cross section measurements in a high-density SEU Hardened SRAM," *IEEE Trans. Nucl. Sci.*, Vol. 52, No. 6, pp. 2125–2131, Dec. 2005.

[24] A. Bogorad, J. Likar, R. Lombardi, S. Stone, and R. Herschitz, "On-Orbit Error Rates of RHBD SRAMs: Comparison of calculation techniques and space environmental models with observed performance," *IEEE Trans. Nucl. Sci.*, Vol. 58, No. 6, pp. 2804-2806, Dec. 2011.

[25] G. I. Zebrev, K. S. Zemtsov, R. G. Useinov, M. S. Gorbunov, V. V. Emeliyanov, A. I. Ozerov, "Multiple Cell Upset Cross Section Uncertainty in Nanoscale Memories: Microdosimetric Approach," *RADECS -2015 Proceedings*, pp. 1-5, 2015.

[26] R. Ladbury, "Statistical Properties of SEE rate Calculation in the Limits of Large and Small Event Counts," *IEEE Trans. Nucl. Sci.*, Vol. 54, No. 6, pp. 2113–2119, Dec. 2007.

[27] R. A. Reed, R. A. Weller, R. D. Schrimpf, M. H. Mendenhall, K. M. Warren, L. W. Massengill, "Implications of nuclear reactions for Single Event Effects Test Methods and Analysis," *IEEE Trans. Nucl. Sci.*, V.53, No.6, pp. 3356-3362, Dec. 2006.

[28] R. A. Weller, R. D. Schrimpf, R. A. Reed, M. H. Mendenhall, K. M. Warren, B. D. Sierawski and L. W. Massengill, "Monte Carlo Simulation of Single Event Effects," *RADECS 2009 Short Course*.

[29] K. M. Warren, B. D. Sierawski, R. A. Reed, R. A. Weller, C. Carmichael, A. Lesea, M. H. Mendenhall, P. E. Dodd, R. D. Schrimpf, L. W. Massengill, Tan Hoang, Hsing Wan, J. L. De Jong, R. Padovani, J. J. Fabula, "Monte-Carlo Based On-Orbit Single Event Upset Rate Prediction for a Radiation Hardened by Design Latch," *IEEE Trans. Nucl. Sci.*, Vol. 54, No. 6, pp. 2419–2425, Dec. 2007.

[30] P. E. Dodd, J. R. Schwank, M. R. Shaneyfelt et al., "Impact of Heavy Ion Energy and Nuclear Interactions on Single-Event Upset and Latchup in Integrated Circuits," *IEEE Trans. Nucl. Sci.*, Vol. 54, No. 6, pp. 2303–2311, Dec. 2007.